\documentclass[preprint,amsmath,amssymb,aps,nofootinbib]{revtex4-1}

\usepackage{graphicx}
\usepackage{epstopdf}
\usepackage{amsmath}
\usepackage{bbold}
\usepackage{times}
\usepackage{float}
\usepackage[h]{esvect}
\usepackage{appendix}
\usepackage{chemarr}
\usepackage{multirow}
\usepackage{natbib}
\usepackage[symbol]{footmisc}

\usepackage{subfig}

\makeatletter

\newcommand\tabcaption{\def\@captype{table}\caption}
\newcommand{\ket}[1]{\left|#1\right\rangle}

\begin{document}

\title{Energy-dependent 3-body loss in 1D Bose gases}

\author{Laura A. Zundel, Joshua M. Wilson, Neel Malvania}
\author{Lin Xia}
\altaffiliation[Present address: ]{Beijing National Laboratory for Condensed Matter Physics, Institute of Physics, Chinese Academy of Sciences, Beijing 100190, China}
\author{Jean-Felix Riou}
\author{David S. Weiss}
\email[email: ]{dsweiss@phys.psu.edu}
\affiliation{ Physics Department,  The Pennsylvania State University, 104 Davey Lab, University Park, PA 16802}

\date{\today}

\begin{abstract}

%We measure the three-body loss in out-of-equilibrium one-dimensional Bose gases and find that the loss rate depends strongly on the average energy of the distribution, and on the transverse oscillator length. Theory predicts that for a strictly one-dimensional system, the collision cross-section depends on both of these quantities.  After adapting this theory to a waveguide, we compare our measured energy dependence to theoretical prediction and see that they agree well. However, our measured dependence on the transverse oscillator length is inconsistent with theoretical prediction.

% We measure the three-body loss in out-of-equilibrium quasi-one-dimensional  Bose gases and find that the loss rate depends strongly on the average energy of the distribution and the transverse confinement. For a strictly one-dimensional system, the theoretical collision cross-section is predicted to depend strongly on the collision energy and the one-dimensional scattering length. After introducing a rolloff to account for the fact that the atoms are quasi-1D, we find that the theory captures the essential features of our data.

% In the absence of a complete theory of loss in this out-of-equilibrium gas, we can approximately reproduce the data with two empirical models of three-body collisions, one based on energy-dependent correlations and the other on a collision energy-dependent loss rate.

%Both models capture the energy-dependence of loss reasonably well, which suggests that they may be complementary ways to view the same physics.

We study the loss of atoms in quantum Newton's cradles (QNCs) with a range of average energies and transverse confinements. We find that the three-body collision rate in one-dimension is strongly energy dependent, as predicted by a strictly 1D theory. We adapt the theory to atoms in waveguides, then using detailed momentum measurements to infer all the collisions that occur, we compare the observed loss to the adapted theory and find that they agree well.

\end{abstract}
%\pacs{34.50.Cx, 34.90.+q, 34.10Jk}

\maketitle

When three ultracold atoms exit a collision as a molecule and an atom, both particles escape typical atom traps.  While these inelastic 3-body collisions are often an atomic density limiting nuisance \cite{lobser2015observation}, they can sometimes be usefully employed. They enable the observation of Efimov triplet states \cite{huckans2009three,ottenstein2008collisional}, they allow detection of triply occupied states in optical lattices \cite{xia2015quantum}, they can cool 1D gases \cite{syassen2008strong,schemmer2018cooling}, and they can probe local three body correlation functions in 3D \cite{burt1997} and 1D Bose gas experiments \cite{tolra2004,haller2011,smirne2007collisional}. When the two-body scattering length exceeds other length scales, as it often does, 3-body inelastic collisions exhibit analytically calculable universal behavior \cite{esry1999recombination,greene2004revised,mehta2007}. The collision rate in 3D is energy-independent \cite{d2004}, but a steep energy dependence has been predicted in 1D \cite{mehta2007}.

In 3D, inelastic 3-body collision rates have been measured in both thermal and quantum degenerate gases \cite{burt1997}, but in 1D they have only been measured in the latter \cite{tolra2004,haller2011} because one needs to reach very low temperatures to satisfy the quasi-1D requirement of being only in the ground state of a waveguide \cite{riou2014}. By making a QNC excitation \cite{kinoshita2006}, which we explain below, we give our 1D gases kinetic energies that exceed the quantum degeneracy energy scale. After the initial oscillations dephase, the atoms' axial motion can be well described semiclassically, so 3-body inelastic collisions involve distinct particles, rather than emerging from overlapping, correlated wavefunctions. Thus, independent of the many-body physics that dominates quantum degenerate gases, we can study the 3-body loss constant itself, $K_3^{1D}$, observing its dramatic cubic dependence on center of mass collision energy, $E_{cm}$ \cite{mehta2007}. Our results are not inconsistent with a predicted sixth order dependence on the 1D scattering length, $a_{1D}$, although our experiment is rather insensitive to this dependence.

We begin our experiments with a BEC of $\sim 5\times10^5$ $^{87}$Rb atoms in the $\ket{F=1,m_F=1}$ state. A 2D blue-detuned optical lattice with wavevector $k=2\pi /772$ nm is ramped up to a lattice depth $V_0=40$$E_R$ in 23 ms, where the recoil energy $E_R=(\hbar k)^2/2m$ and $m$ is the mass of $^{87}$Rb. We thus create an array of $\sim 3500$ tubes with minimal tunneling between tubes. After loading the atoms into the 2D lattice, the atom cloud has a Thomas-Fermi radius of 13.0 $\mu$m and $4.9\times10^5$ atoms. We also create somewhat lower density distributions by lengthening the ramp up time to 70 ms, resulting in a distribution with a Thomas-Fermi radius of 13.9 $\mu$m and $4.5\times10^5$ atoms. Axial confinement is provided by a red-detuned optical dipole trap of depth $U_o=10$$E_R$, chosen to be less than half the energy splitting  between the transverse ground and second excited states. This condition ensures that even the most highly axially excited atoms do not have enough  energy to excite to higher transverse states via binary collisions. Also, most atoms that are transversely excited due to spontaneous emission are lost when they collide among themselves and vibrationally de-excite \cite{riou2014}.

%[Move to Methods, which I think is Supplemental for NP] We use a measurement of the total energy of the equilibrium 1D Bose gases to calibrate our density. We first shut off the dipole trap and let the atoms expand in the 2D lattice for 15 ms. We then absorption image the atoms and extract their number and expansion energy \cite{kinoshita2004}. We assume a Thomas-Fermi distribution of atoms among the tubes, $N_{tube}=N_{max}(1-(x/R_{xy})^2-(y/R_{xy})^2)^{3/2}$, where $N_{max}=5N\pi/(2k^2R_{xy}^2)$, and $R_{xy}$ is the transverse radius \cite{dunjko2001}. Using the Lieb-Liniger model and the local density approximation for each tube \cite{dunjko2001}, we find the $R_{xy}$ for which the theoretical expansion energy best matches the observation ($R_{xy}=13.90 +0.20/-0.16$ $\mu$m).

\begin{figure}[h]
\begin{center}
\includegraphics[width=5.5 in]{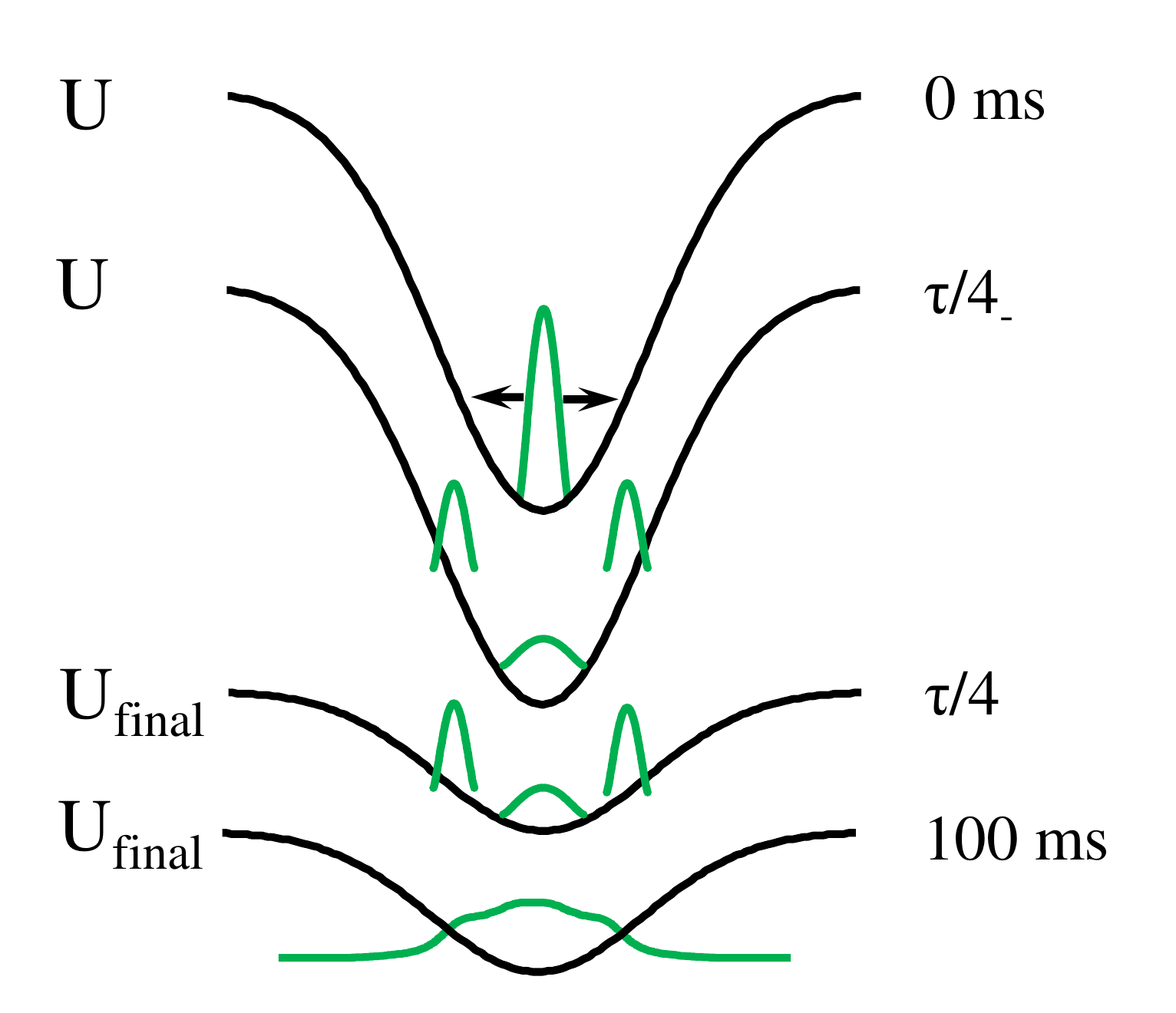}
\caption{(Color Online) Axial trap sequence used to change the average energy of the atoms. At $t=0$ ms most of the atoms are diffracted and the axial trap
depth is  immediately increased from $U_o$ to $U_i$. When the atoms reach their turning point a quarter-cycle later the axial trap is decreased to $U_{final}=\{9.3, 10, 11, 12\}$$E_R$ for $V_{latt}=\{36, 40, 45, 50\}$$E_R$ respectively. The atoms then evolve for 200 ms, by which time the distributions are fully dephased. }
\label{YAGtiming}
\end{center}
\end{figure}

We take the 1D gases out of equilibrium by setting up QNCs \cite{kinoshita2006} of varying average energy, $\overline{E_o}$. The QNC is started by applying a sequence of two standing wave pulses along the axis of the tubes, made from beams with wavenumber $\sim k$ and intensity 18 W/cm$^2$ \cite{wu2005}. Each pulse is 23 $\mu s$ long and there is a 33 $\mu s$ pause between pulses. Most atoms end up in a superposition of $\pm 2\hbar k$ momentum states, although due to interactions $\sim 30\%$ of the atoms remain near zero momentum. Immediately after diffraction, the lattice and axial depths are suddenly ramped to $V_{latt}$ and $U_i$ respectively, where $V_{latt}$ ranges from 36$E_R$ to 50$E_R$.  The atoms are allowed to expand in the deeper trap for a quarter of the oscillation period of $U_i$ (see Fig.\ref{YAGtiming}). At their turning point, we  remove part of their mechanical energy by suddenly decreasing the axial depth to $U_{final}$, which is chosen to be less than half the energy splitting between the transverse ground and second excited states of the corresponding $V_{latt}$. We then wait for approximately $20\tau$, where $\tau=10$ ms is the typical axial oscillation period in the final trap. During this time in the Gaussian, and thus significantly anharmonic axial trap, the atoms' motion dephases within each tube and the density distribution ceases to change on the $\tau$ timescale. We continue to let the atoms evolve for a time $t_{ev}$. At $t_{ev}$ we turn off the dipole trap and adiabatically ramp down the lattice to 2.5$E_R$ in 0.14 ms in order to remove most of the transverse trapping energy and residual atom interaction energy. We then suddenly complete the turn-off of the lattice light and let the atoms expand for 15 ms time-of-flight.

\begin{figure}[h]
\begin{center}
\includegraphics[width=5 in]{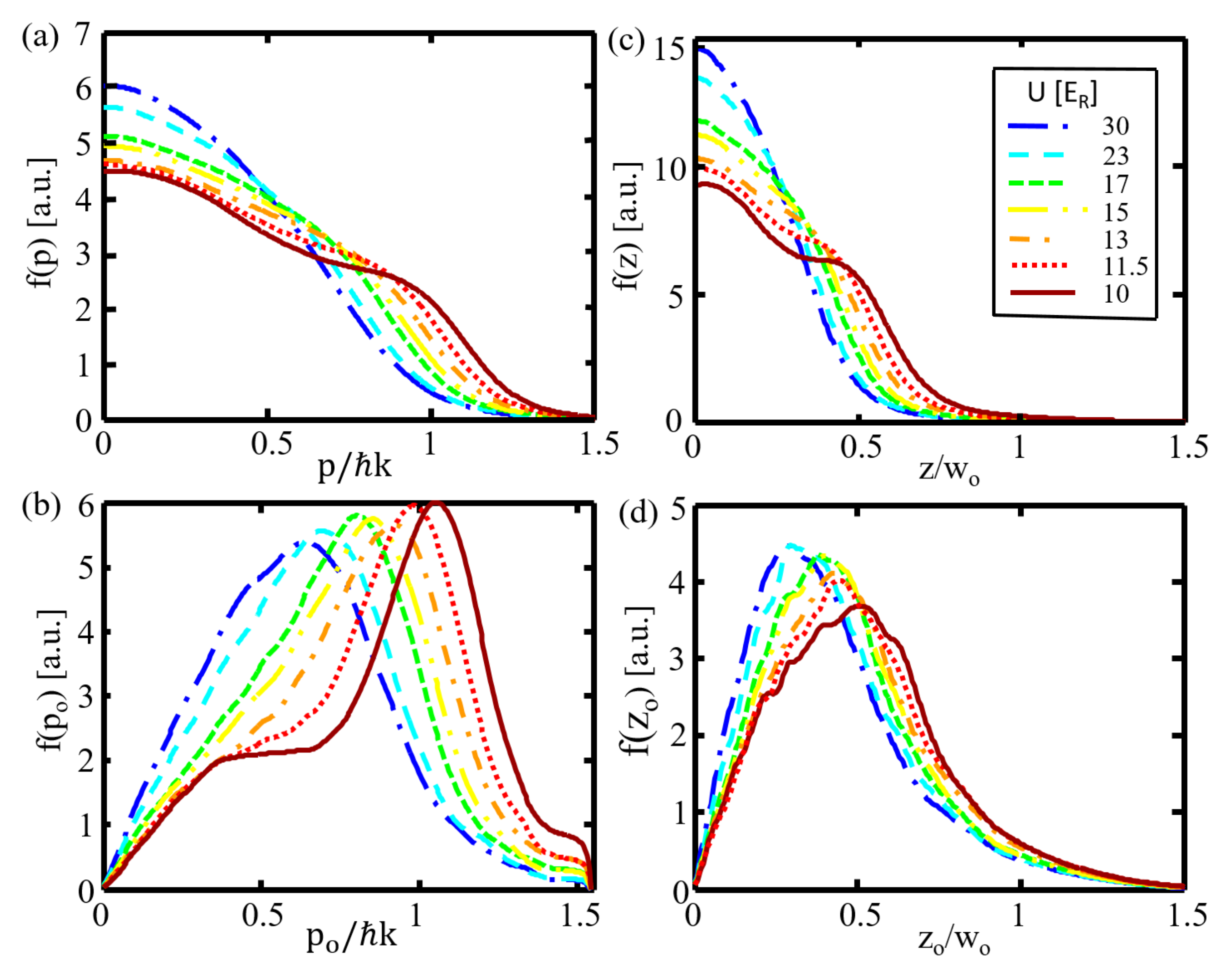}
\caption{(Color Online) Normalized 1D trap distributions, for various QNC energies and a $45$$ E_R$ lattice confinement. (a) $f(p)$. (b) $f(p_o)$. (c) $f(z)$. (d) $f(z_o)$. The curves for $f(p)$ and $f(z)$ are normalized to 0.5, while the $f(p_o)$ and $f(z_o)$ distributions are normalized to 1. In (c) and (d) $w_0=42.6$ $\mu$m is the beam waist of the axial dipole trap.}
\label{distributions}
\end{center}
\end{figure}

We take an absorption image at each $t_{ev}$, integrating transversely to get the axial momentum distribution, $f(p)$, and integrating the whole image to find the total atom number.  The initial $f(p)$ for various $\overline{E_o}$ are shown in Fig. \ref{distributions}a. Because the atoms have more than 20 times as much mechanical energy as interaction energy, their motion can be well described semi-classically \cite{riou2014}. Since loss depends on the axial spatial distribution, $f(z)$, and the collision energy of the particles, we derive both $f(z)$ and the distribution of peak momenta, $f(p_o)$, from $f(p)$  via energy conservation as follows. The mechanical energy of each atom is $E_o=p_o^2/2m=p^2/2m+U(z)$, where $p_o$ is the amplitude of the atom's momentum oscillation and $U(z)$, a Gaussian function, is the potential energy at the axial position $z$. We extract from $f(p)$ the distribution of peak momenta $f(p_o)$ using the relation $f(p)=\int_0^{p_o} G(p,p_o)f(p_o) dp_o$, where $G(p,p_o)$ is the probability that an atom with momentum amplitude $p_o$ has momentum $p$ \cite{riou2014}; the results are shown in Fig. \ref{distributions}b. To derive $f(z)$, we use conservation of energy to convert $f(p_o)$ to $f(z_o)$, the axial spatial amplitude distribution (Fig. \ref{distributions}d). Then, using a procedure like the inverse of the transformation from $f(p_o)$ to $f(p)$, we calculate $f(z)$ from $f(z_o)$; it is plotted in Fig. \ref{distributions}c.

\begin{figure}[h!]
\centering
\includegraphics[width=5.5 in]{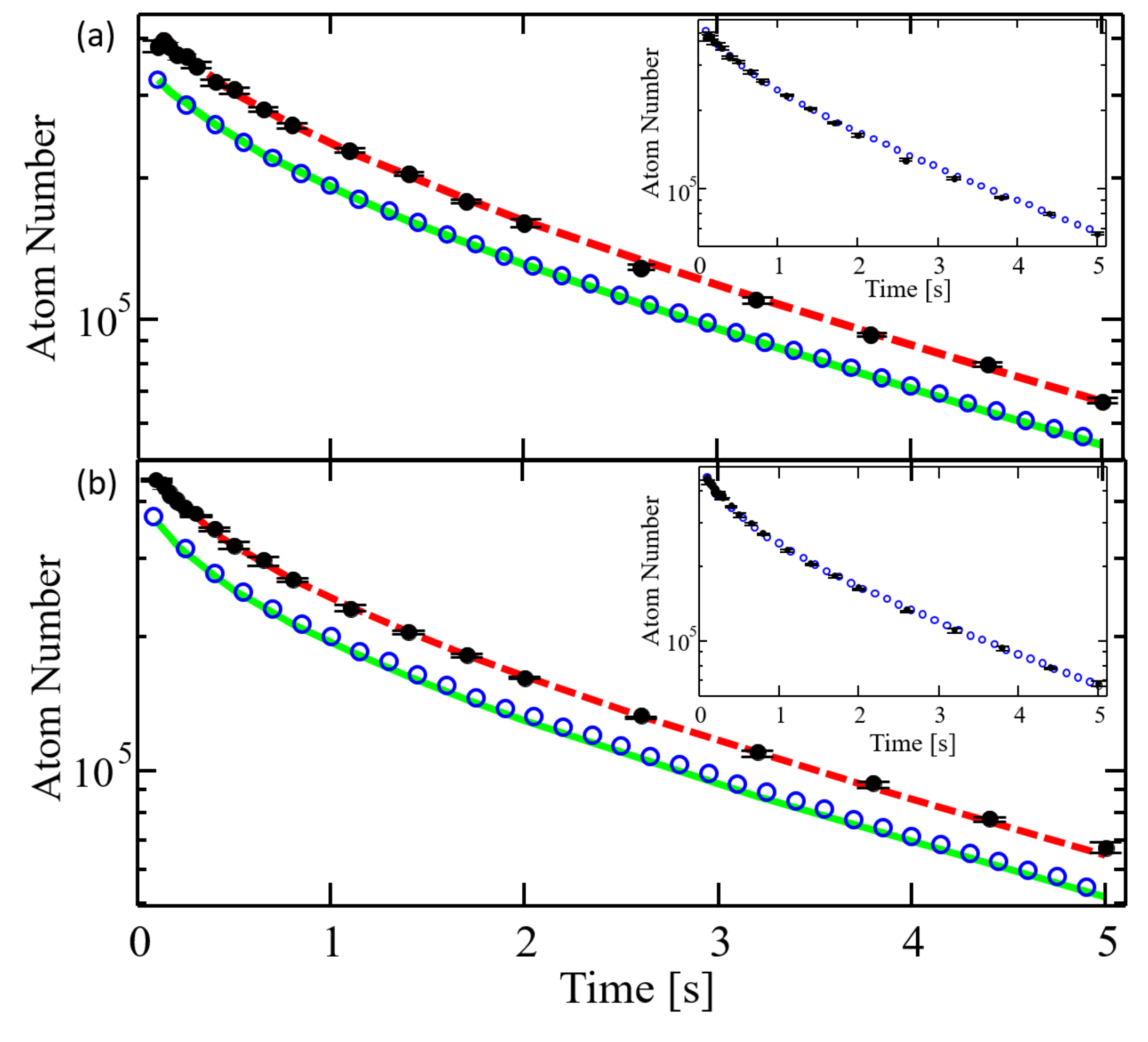}
\caption{(Color Online) Total atom number as a function of holding time in a (a) lower and (b) higher density QNC distribution with $\overline E_o=3.05$$E_R$ in a $50$$E_R$ lattice depth. Black circles are the average of six data points; the standard deviation of these points are smaller than the marker size. The red dashed lines are the fits of the experimental data curves to Eq. \ref{eq:K3eff}. The insets show the direct comparison between the experimental data curves and the theoretical curves (open blue circles) derived from the global fit of the loss model (Eq. \ref{eq:E3loss}) to the data. In order to make many such comparison on the same plot (see Fig. \ref{K3Fits}), we fit the theoretical curves to Eq. \ref{eq:K3eff} (solid green lines in the main figures). To make these curves more visible, they have been offset by a factor of 0.8.}
\label{LossFigure}
\end{figure}

\begin{figure}[h]
\begin{center}
\includegraphics[width=5.5 in]{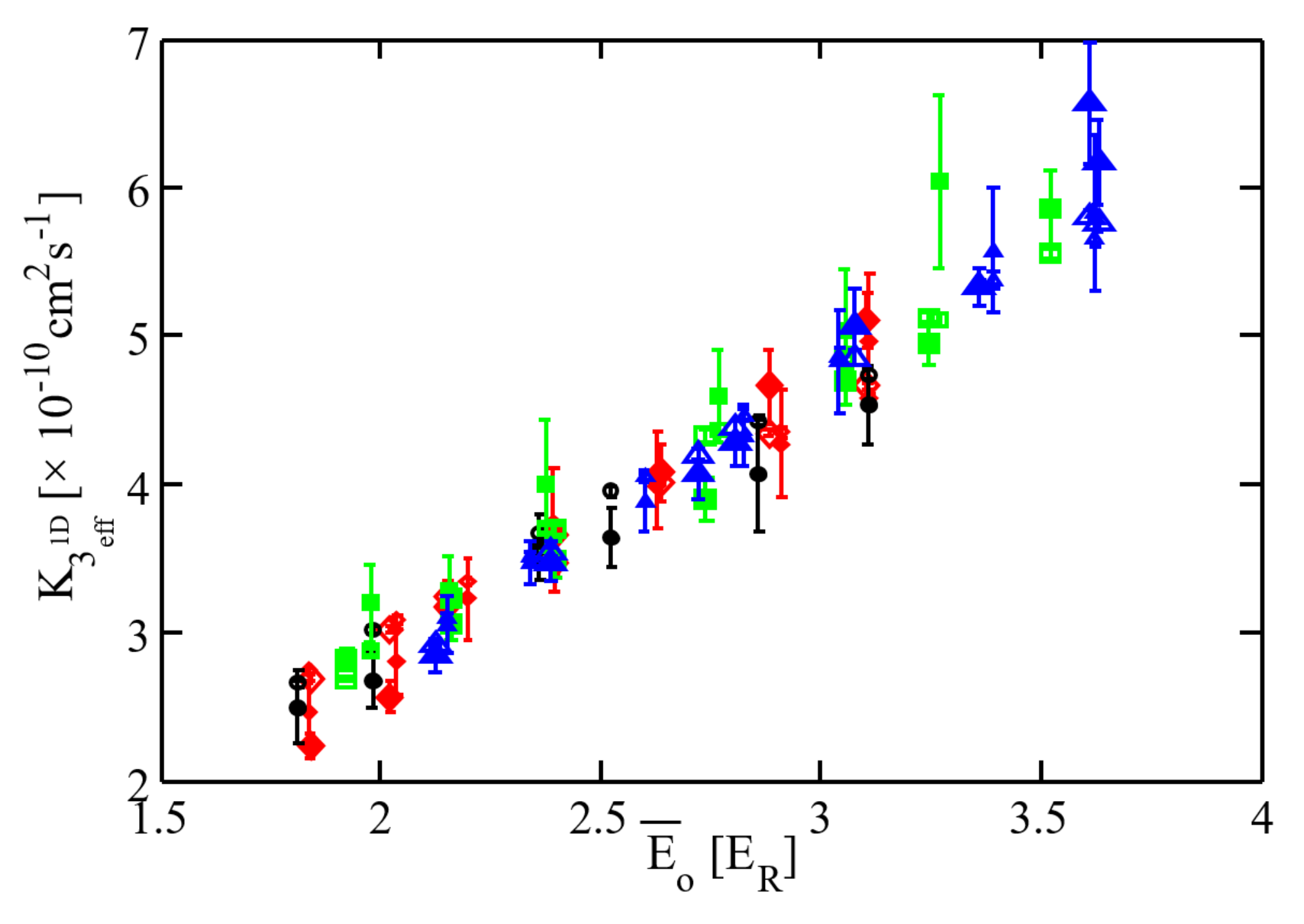}
\caption{(Color Online) Experimental and theoretical $K^{1D}_{3_{eff}}$. The experimental points are the solid symbols (larger for higher density and smaller for lower density)  shown as red diamonds, black circles, green squares, and blue triangles for $36$, $40$, $45$, and $50$$E_R$ lattice depths respectively. We obtain the experimental points  by fitting the experimental loss curves to Eq. \ref{eq:K3eff}. $K^{1D}_{3_{eff}}$ increases with energy at all lattice depths, but does not depend on the lattice depth. We obtain the theoretical points (hollow symbols) by fitting the decay curves derived from our detailed loss model  (Eq. \ref{eq:E3loss}) to Eq. \ref{eq:K3eff}.}
\label{K3Fits}
\end{center}
\end{figure}

We measure loss curves like those shown by the black circles in Fig. \ref{LossFigure} for each initial $\overline{E_o}$. To qualitatively see how the loss depends on energy, we assume each loss curve can be described by an effective three-body loss coefficient, $K^{1D}_{3_{eff}}$, and fit each curve to:
\begin{equation}
\label{eq:K3eff}
\frac{dN}{dt}=-K_1N-K^{1D}_{3_{eff}}N^3\int f(z)^3dz
\end{equation}
where  $K^{1D}_{3_{eff}}$, $K_1$ and the initial atom number, $N_o$, are fit parameters.  $K_1$ is mostly due to lattice spontaneous emission, and it depends on $V_{latt}$ and $U_{final}$, so we constrain all curves with the same $V_{latt}$ and $U_{final}$ to have the same $K_1$ \cite{riou2014}. These fits are shown by the red dashed lines in Fig. \ref{LossFigure}. We plot $K^{1D}_{3_{eff}}$ as a function of $\overline{E_o}$, shown in Fig. \ref{K3Fits} by the solid red diamonds, black circles, green squares, and blue triangles corresponding to $V_{latt}=36$, $40$, $45$, and $50$$E_R$. The larger and smaller symbols correspond to higher and lower density distributions respectively. If the collisions were energy-independent, the plot would be a horizontal line. Instead, the data shows that distributions with higher $\overline{E_o}$ have a higher $K^{1D}_{3_{eff}}$. The strong energy dependence we measure arises even though there is considerable averaging inherent to the broad $f(p)$ distributions in a QNC. The data in Fig. \ref{K3Fits} also shows that the loss is independent of lattice depth, in apparent contradiction of the 1D semi-classical model.

\begin{figure}[h]
\begin{center}
\includegraphics[width=3.7 in]{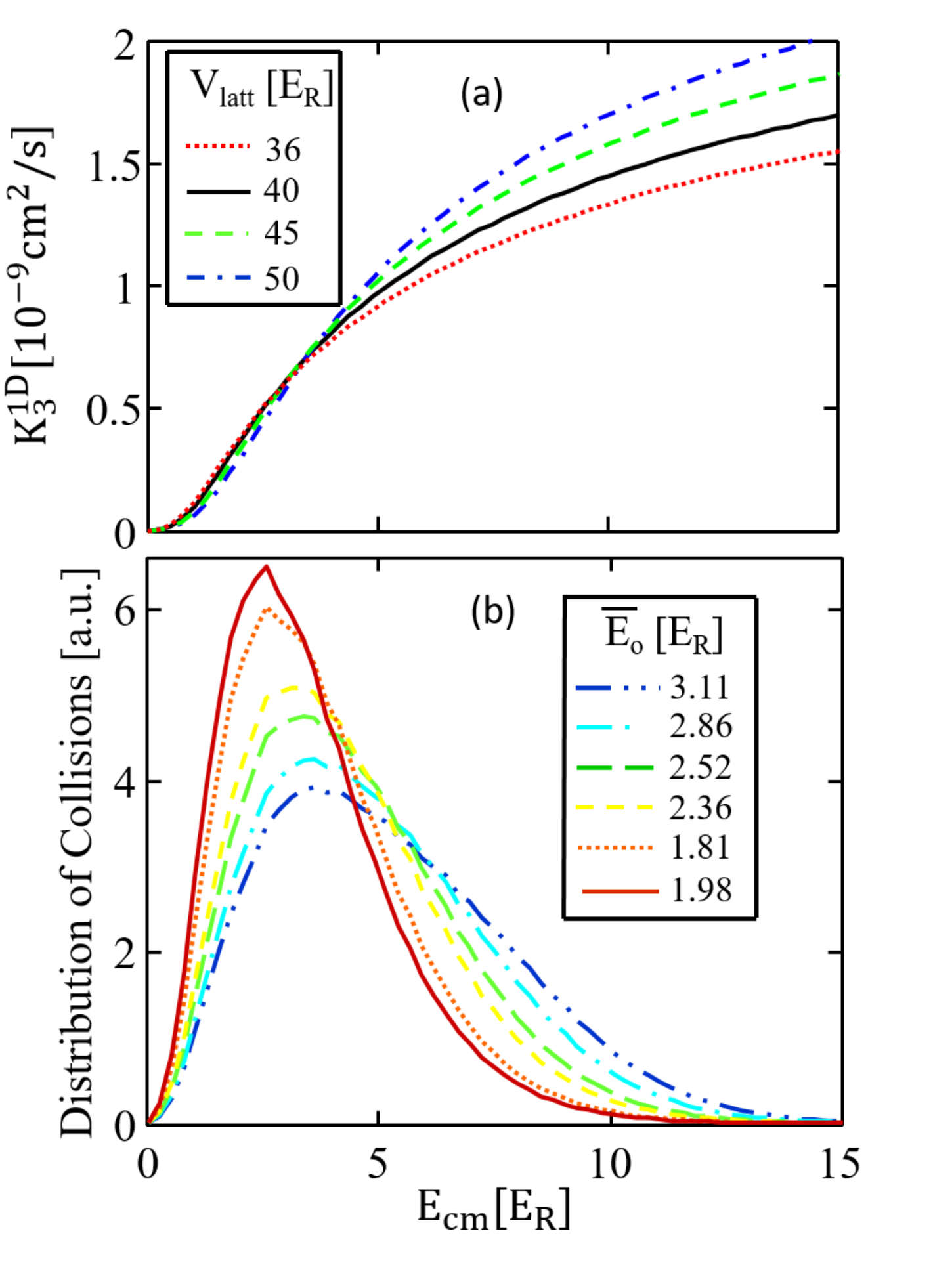}
\caption{(Color Online) (a) Plot of $K_3^{1D}$ as a function of the center of mass collision energy for atoms in a $36$, $40$, $45$, and $50$$E_R$ lattice. A higher maximum value corresponds to a deeper lattice depth.  (b) Plot of the distribution of collisions in the $40$$E_R$ lattice as a function of the collision energy for different $\overline{E_o}$.}
\label{K31D}
\end{center}
\end{figure}

To model the energy dependence of loss in a QNC, we start with the strictly 1D semi-classical model of three-body collisions \cite{mehta2007} which gives $K_3^{1D}=Ca_{1D}^6E_{cm}^3$, where $K_3^{1D}$ is the three-body collision cross section in 1D and $C$ is a proportionality constant that depends on the atom. The basic adaptation of 1D theory to atoms in a waveguide has long been known, $a_{1D}\approx-2a_{\perp}^2/a_{3D}$ \cite{olshanii1998}. Since we expect loss to only be suppressed in 1D compared to 3D, not enhanced, we expect the loss rate at high $E_{cm}$ to saturate at the thermal 3D loss rate, $6\times K_3$, where $K_3$ is the three-body collision cross section in a 3D quantum degenerate gas \cite{burt1997}. We therefore introduce $E_l$, a collision energy to  characterize this saturation. Since the transverse vibrational excitation energy is characteristic of the crossover to the 3D, we assume that $E_l$ scales with it, i.e., $E_l=2 C_2\hbar^2/ma_{\perp}^2$, where $C_2$ is a constant to be empirically determined. Perhaps the simplest function that smoothly meets our criteria at low and high energies with a variable crossover energy is:
\begin{equation}
\label{eq:K31D}
K_3^{1D}(E_{cm})=C'a_{\perp}^{12}E_{cm}^3\bigg(\frac{C'a_{\perp}^{12}E_{cm}^3}{K_{3_{max}}^{1D}(\frac{E_{cm}}{E_{cm}+E_l})}+1\bigg)^{-1}
\end{equation}
where $K_{3_{max}}^{1D}=6\times K_3/3\pi^2a_{\perp}^4$ \cite{tolra2004}. We have also tried other ways to roll off $K_3^{1D}$ at high energy, and we find that the fit quality is not very sensitive to the form of the roll off \footnote{See Supplemental Material for more information on the sensitivity to the roll off.}. % and Fig (XX insert fig reference) shows the distribution of collisions as a function of the collision energy for the data taken in the $40E_r$ lattice.

 We calculate the loss rate for all possible collisions by first discretizing the $f(z_o)$ distribution. We convert each $f(z_o)$ segment into corresponding spatial distributions, %that depend on both $z$ and $z_o$
 $f(z,z_o)$, then divide the total spatial distribution $f(z)=\sum_{z_o}f(z,z_o)$  into small segments of width $\Delta z$. Each atom within a $\Delta z$ segment has a defined momentum that can be calculated from $z$ and $z_o$. These momenta are then used to calculate $E_{cm}$ for all momentum combinations, including all direction combinations. From the $E_{cm}$'s we use Eq. \ref{eq:K31D} to calculate the  corresponding $K_3^{1D}$'s, which are then used to calculate the three body loss within each $\Delta z$. This loss is given by
\begin{equation}
\label{eq:E3loss}
\frac{dN}{dt}=-K_1N-N^3\sum_{z_{o_i}=z_{o_1}}^{z_{o_f}}\sum_{z_{o_j}=z_{o_i}}^{z_{o_f}}\sum_{z_{o_k}=z_{o_j}}^{z_{o_f}}K_3^{1D}(E_{cm})\int_z^{z+\Delta z}f(z',z_{o_i},t)f(z',z_{o_j},t)f(z',z_{o_k},t)dz'
\end{equation}
The lowest energy $z_o$ group is $z_{o_1}$ and the highest is $z_{o_f}$. To account for all collisions in $\Delta z$ without double counting, collisions of particles with $z_{o_i}$, $z_{o_j}$ and $z_{o_k}$ are weighted by their relative probability.
%Collisions between atoms with different $z_o$ are twice as likely as collisions where two atoms have the same $z_o$, and six times more likely than collisions where all three atoms have the same $z_o$.
We use the measured $f(z)$ distributions as time evolves.

We use the $K_1$'s and $N_o$'s obtained from the fit to Eq. \ref{eq:K3eff} to globally fit Eq. \ref{eq:E3loss} to all of the loss data curves  with $C'$ and $C_2$ as free parameters, while $K_3$ is fixed to the weighted average of previous measurements \cite{burt1997,tolra2004}, $K_3=7.1\times10^{-30}$ cm$^6$/s. Because the fit quality is  sensitive to the ratio of cloud sizes for the higher and lower densities, we fit using the measured size of the lower density cloud and allow the higher density cloud size to be a free parameter, restricting its value to be within its measurement uncertainty.  We obtain the fit parameters $C'=4.2(+0.18/-0.16)\times10^{142}$  cm$^{-10}$J$^{-3}$s$^{-1}$ and $C_2=28(+1.6/-3.1) \times10^{-2}$. The bounds on the fit parameters are set by the uncertainty in the lower density cloud size. The blue circles in Fig. \ref{LossFigure} show examples of the theoretical curves generated by this elaborate fit based on keeping track of all 3-body collisions.

Figure \ref{K31D}a shows $K_3^{1D}(E_{cm})$ at all lattice depths using these fit parameters. Fig. \ref{K31D}b shows the distribution of inelastic collisions as a function of $E_{cm}$ for various values of $\overline{E_o}$ in a 40$E_R$ lattice. We calculate this distribution by summing together the collisions that enter into Eq. \ref{eq:E3loss} and sorting them by $E_{cm}$.
 The red dotted 36$E_R$ line in Fig. \ref{K31D}a has a somewhat higher $K_3^{1D}$ for $E_{cm}<4$$E_R$ where there are many collisions, while the blue dash-dotted 50$E_R$ line has a much higher $K_3^{1D}$ for $E_{cm}>4$$E_R$ where there are fewer collisions. The net effect is very weak dependence on lattice depth.

It is hard to visualize how the model fits the data by directly comparing the many theoretical curves to their corresponding experimental data curves (insets in Fig. \ref{LossFigure}). Instead,  we fit each of the theoretical curves (like the blue circles in Fig. \ref{LossFigure}) to Eq. \ref{eq:K3eff}, using $K^{1D}_{3_{eff}}$ as a free parameter just as we did for the experimental data curves. These fits are shown by the green lines in Fig. \ref{LossFigure}. The $K^{1D}_{3_{eff}}$'s obtained by fitting the theoretical curves derived from the model (hollow symbols in Fig. \ref{K3Fits}) can then be compared to the $K_{3_{eff}}^{1D}$'s obtained by fitting the experimental loss (solid symbols in Fig. \ref{K3Fits}). The reduced chi-square value between the $K^{1D}_{3_{eff}}$'s for the theory and the data is $2.08$. The model captures the observed increase in $K^{1D}_{3_{eff}}$ with energy at all $V_{latt}$, and the approximate independence of $K^{1D}_{3_{eff}}$ on $V_{latt}$ even though the loss model is $a_{\perp}$-dependent.   

%The actual shape of the roll off assumed in Eq. \ref{eq:K31D} is not theoretically known. To see how sensitive we are to the shape of the roll off, we have fit a variety of $K_3^{1D}(E_{cm})$ functional forms (see Supplemental Material). By comparing the $K_3^{1D}$ curves and their respective fit qualities we are able to constrain a region where the best fits lie. The $K_3^{1D}$ used here bounds this region, indicating that the shape we assumed for the roll off agrees well with our loss measurements.

Studies of low temperature 1D gases have used a rather orthogonal way to look at 3-body loss, based on local correlations, $g_3(0)$, with no $E_{cm}$ dependence. As far as we know, the relationship between loss suppression in 1D by reduced $g_3(0)$ and by energy dependent $K_3^{1D}$ has not been explored theoretically. The latter cannot readily be applied in equilibrium, where there is no possible semi-classical approximation. The former is problematic for a QNC where the system is not in thermal equilibrium and correlations must change dynamically as the atoms move \cite{kheruntsyan2003pair}. Applying both approaches at the same time would constitute double counting that is inconsistent with observations. Loss suppression by reduced $g_3(0)$ and energy dependent $K_3^{1D}$ may be complementary ways of describing much of the same physics. More theory is needed.

In conclusion, we have measured the three-body loss rates in out-of-equilibrium 1D Bose gases with different average energies and transverse confinements. We find that they strongly depend on the energy of the colliding atoms, as a purely 1D theory predicts \cite{mehta2007}.   Still, a rigorous theory of 3-body inelastic collisions in waveguides is needed. Besides testing such a theory, our results will be important in studies of thermalization in out-of-equilibrium 1D gases \cite{kinoshita2006, gring2012}.

\section*{Acknowledgements}
This works was supported by the Army Research Office Grant No. W911NF-16-1-0031 and NSF Grant No. PHYS-1707576.

%\section*{Author Contributions}
A%ll authors made significant intellectual contributions to this work. L.Z. did all the modeling of the experiment, L.Z., J.W., N.M. and L.X. took the data, and D.W. oversaw all aspects of the experiment.

%\section*{Competing financial interests}
%The authors declare no competing financial interests.

\vspace{1 cm}

\bibliographystyle{apsrev4-1}
\bibliography{bibFile}

\end{document}

\title{Energy-dependent 3-body loss in 1D Bose gases \\
	\large{Supplementary Material}
}

\author{Laura A. Zundel, Joshua M. Wilson, Neel Malvania,  Lin Xia,  Jean-Felix Riou and David S. Weiss}
\affiliation{ Physics Department,  The Pennsylvania State University, 104 Davey Lab, University Park, PA 16802}

\date{\today}

\section{Fit Sensitivity to the shape of $K_3^{1D}$}
The low and high energy limits of the functional form of $K_3^{1D}(E_{cm})$ are theoretically known, but the roll off transition between these two limits was postulated for its simplicity. It is reasonable, then, to ask how sensitive our results are to the exact form of this roll off. In this supplemental material we fit several different parameters and functional forms for $K_3^{1D}$. We first test our sensitivity to the measured value of $K_3$. We then fit several different functional forms of $K_3^{1D}$ to the data including a function with a faster roll off, an $a_{\perp}$-independent function, and a function with linear dependence on $E_{cm}$. We compare the fit qualities for each of these functions to empirically constrain the range of shapes of $K_3^{1D}(E_{cm})$ that are consistent with our observations.

\begin{figure}[h]
\begin{center}
\includegraphics[width=5.5 in]{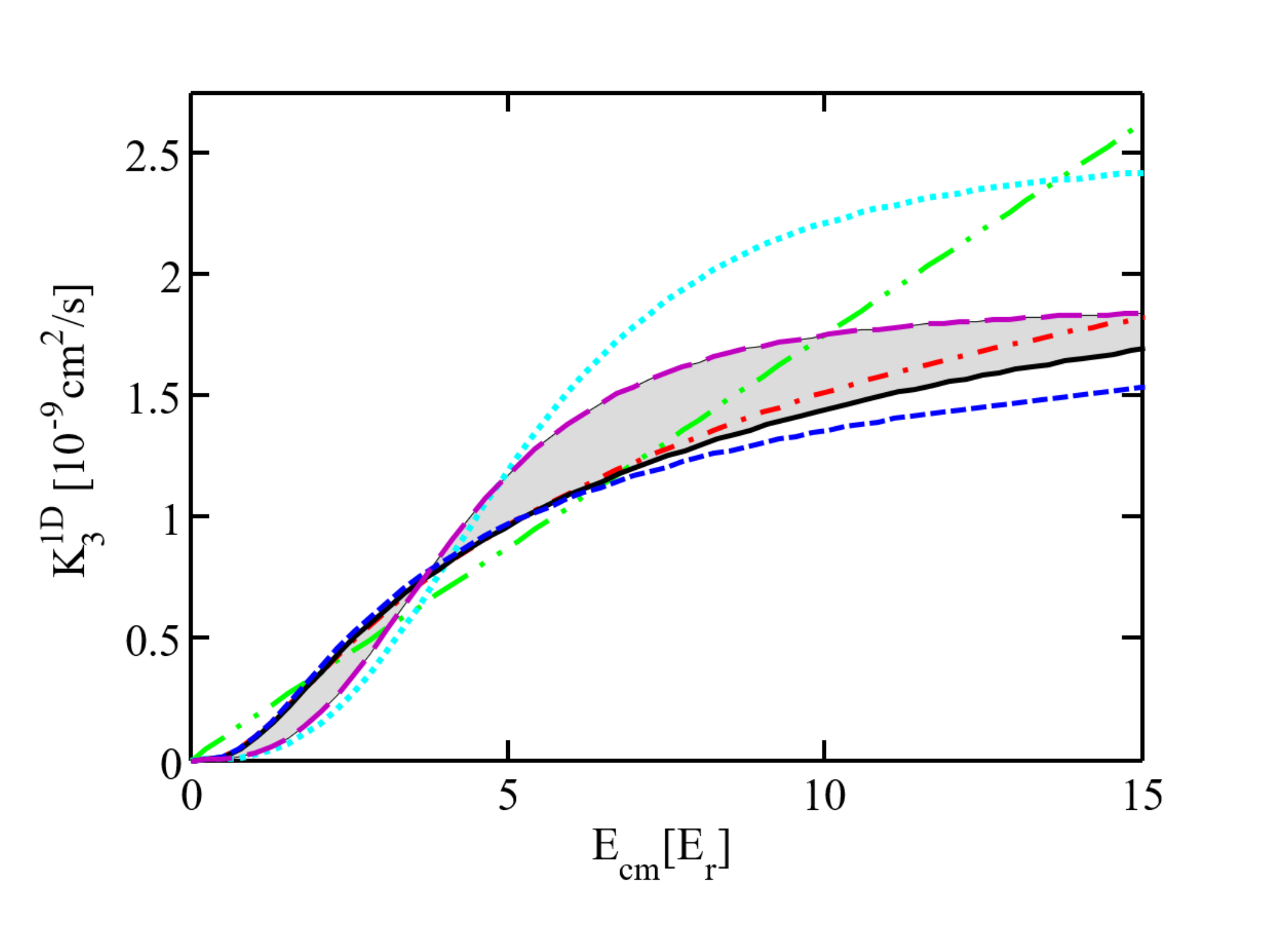}
\caption{(Color Online) $K_3^{1D}$ curves for $K_{3_{\text{base}}}^{1D}$ (solid black line), $K_{3_{\text{lb}}}^{1D}$ fit (dashed blue line), $K_{3_{\text{ub}}}^{1D}$ fit (red dash-dotted line), $K_{3_{alt}}^{1D}$ (dotted light blue line), $K_{3_{\text{LI}}}^{1D}$ (long dashed purple line), and $K_{3_{\text{lin}}}^{1D}$ (dash-dot-dotted green line). The qualities of these fits are $\chi^2=\{2.08, 2.55, 1.98, 3.7, 1.8, 2.4\}$ respectively. The gray shaded region indicates where the optimal form of $K_3^{1D}$ likely lies.}
\label{fig:Supplemental1}
\end{center}
\end{figure}

 First we test how sensitive the fit is to the exact value of $K_3$. In the main text $K_3$ is fixed to the weighted average of the measured values, but there is some uncertainty associated with those measurements. Here, we refit the data again with $K_3$ fixed to the lower and upper bounds of that uncertainty, The parameters of these fits are shown in Table \ref{tab:parameters}.  Figure \ref{fig:Supplemental1} compares $K_3^{1D}$ generated by these fits for the $40$ $E_R$ lattice (blue dashed and red dash-dotted lines) to the $K_3^{1D}$ curve from the original fit described in the main text  (solid black line), which we will call $K_{3_{\text{base}}}^{1D}$ in this supplement. The fit quality of $K_{3_{\text{lb}}}^{1D}$ is somewhat worse than the original fit, $\chi^2=2.55$, while that of $K_{3_{\text{ub}}}^{1D}$ is slightly better, $\chi^2=1.98$. This suggests that the maximum value of the optimal fit is closer to the upper boundary of $K_3$.

The simple form for the high energy roll off that we use in the paper is relatively slow. For example, $K_{3_{\text{base}}}^{1D}$ only reaches 78\% of its high energy value at $E_{cm}=26$ $E_R$ (our highest collision energy). To test our sensitivity to the form of the roll off, we have constructed an alternative $K_3^{1D}$ function that is much sharper by cubing all the energy terms in the roll off function,
\begin{equation}
\label{eq:El3}
K_{3_{alt}}^{1D}=C'a_{\perp}^{12}E_{cm}^3\bigg(\frac{C'a_{\perp}^{12}E_{cm}^3}{K_{3_{max}}^{1D}(\frac{E_{cm}^3}{E_{cm}^3+E_l^3})}+1\bigg)^{-1}.
\end{equation}
We globally fit our data using $K_{3_{alt}}^{1D}$ and taking $C'$  and $C_2$ as free parameters while $K_3$ is again fixed to the weighted average of the measured values. The fit values of $C'$ and $C_2$ are shown in Table \ref{tab:parameters} and the corresponding $K_3^{1D}$ curve is shown by the dotted light blue line in Fig. \ref{fig:Supplemental1}. The reduced chi square for this alternative fit function is $\chi^2=3.7$, significantly worse than the original fit.

\newcolumntype{P}[1]{>{\centering\arraybackslash}p{#1}}
\begin{table}
\begin{center}
\begin{tabular}{|c|P{1.5cm} P{1.5cm} P{1.5cm} P{1.5cm}|}
\hline
Fit Function & $K_{3_{\text{base}}}^{1D}$ & $K_{3_{\text{lb}}}^{1D}$ & $K_{3_{\text{ub}}}^{1D}$ & $K_{3_{\text{alt}}}^{1D}$\\
\hline
$C'$ \big($\times10^{142}$\big)\big[cm$^{-10}$J$^{-3}$s$^{-1}$\big] & 4.2 & 3.9 & 3.8 & 2.0\\
$C_2$ \big($\times10^{-2}$\big) & 28 & 19 & 38 & 18\\
$\chi^2$ & 2.08 & 2.55 & 1.98 & 3.7 \\
\hline
\end{tabular}
\end{center}
\caption{Fit parameters and $\chi^2$'s for the fit functions $K_{3_{\text{base}}}^{1D}$, $K_{3_{\text{lb}}}^{1D}$, $K_{3_{\text{ub}}}^{1D}$, and $K_{3_{\text{alt}}}$. The parameters of $K_{3_{\text{LI}}}^{1D}$ and $K_{3_{\text{lin}}}^{1D}$ are not included because they lack physical significance.}
\label{tab:parameters}
\end{table}

The $V_{latt}$ independence of $K_{3_{eff}}^{1D}$ naively suggests that 3-body loss in 1D is, contrary to theory, independent of $a_{\perp}$. We construct a lattice-independent form of $K_3^{1D}$ by removing all $a_{\perp}$ terms, as well as $E_l$ since the premise for its inclusion is that the crossover to 3D is dependent on the transverse vibrational energy. These modifications give
 \begin{equation}
 \label{eq:lattIndependent} 
 K_{3_{\text{LI}}}^{1D}=C_{\text{LI}}E_{cm}^3\bigg(\frac{C_{\text{LI}}E_{cm}^3}{K_{3_{max}}^{1D}}+1\bigg)^{-1}.
 \end{equation}
To keep this model completely lattice independent we need to make the unphysical assumption that $K_{3_{max}}^{1D}$ is the same for all lattice depths, which is inconsistent with a shared $K_{3_{max}}^{3D}$ value. This prevents us from using the measured $K_3$ value for $K_{3_{max}}$, so we instead fit Eq. \ref{eq:lattIndependent} with both $C_{\text{LI}}$ and $K_{3_{max}}$ as free parameters. The result is shown as the long dashed purple curve in Fig. \ref{fig:Supplemental1}. The fit quality is $\chi^2=1.8$, somewhat better than the original fit. Although the lattice-independent model is unphysical, the quality of this fit highlights the fact that our experiment is insensitive to the $a_{\perp}$ dependence of 3-body loss.

Finally, we test to see how sensitive we are to the $E_{cm}^3$ dependence predicted by theory. Given that $K_{3_{eff}}^{1D}$ scales approximately linearly with $\overline{E_o}$ (see Fig. 4), we fit using a linear functional form for $K_3^{1D}$:
\begin{equation}
K_{3_{\text{lin}}}^{1D}=C_{\text{lin}}E_{cm},
\end{equation}
where $C_{\text{lin}}$ is a free parameter. Like ``$K_{3_{\text{LI}}}^{1D}$'' this functional form is not physical, but is useful for determining how sensitive our experiment is to the functional form of $K_3^{1D}$. The fit result is shown by the dash-dot-dotted green line in Fig. \ref{fig:Supplemental1}. For this fit, $\chi^2=2.4$. That we are not very sensitive to how slowly $K_3^{1D}$ rises with $E_{cm}$ at low energies is not surprising in that, as is clear in Fig. 5b, only a small fraction of collisions occur below $\sim1$ $E_R$ where $K_{3_{\text{lin}}}^{1D}$ and $K_3^{1D}$ diverge.

The best fits lie within the shaded region between the black and purple lines in Fig. \ref{fig:Supplemental1}. All the functional forms of $K_3^{1D}(E_{cm})$ we have tried have a positive slope at $E_{cm}=4$ $E_R$ where they pass through $\sim0.75\times10^{-9}$ cm$^2$/s. The best fits start rising more slowly than linear at small $E_{cm}$, and flatten out in the range above $E_{cm}\approx7$ $E_R$. Clearly more theoretical work is needed so that this data can be fit to the correct function, but the ad hoc adaptation of the strictly 1D result that we have used seems to be basically consistent with our observations.

% At low collision energies, where the $E_{cm}^3$ dependence dominates, the different $K_3^{1D}$ curves end up being very similar. At high collision energies the curves diverge and approach maximum values that differ by $\sim20 \%$. There is a reduced chi squared cost of this divergence, illustrating some sensitivity to the high energy part of the curve, but the cost is fairly small: $\chi^2_R=1.28$ for the larger $C$, and $\chi^2_R=1.12$ for the smaller $C$, compared to $\chi^2_R=1.19$ for the best fit. Clearly we are most sensitive to the sharp increase in the collision rate at low energies, compared to how the curve rolls off at high energies. These relative sensitivities can be understood from Figure 5b, which shows that there are relatively few collisions at energies higher than the roll off.

%\begin{figure}[h]
%\begin{center}
%\includegraphics[width=3.7 in]{Supplemental2}
%\caption{ Comparison of $K_3^{1D}(E_{cm})$ for Eq. 2 in the main text (black solid line) and Eq. \ref{eq:El3} (red dashed line). The global fit value of $C$ is approximately the same for both curves. Although the shape of the rolloff differs between the two curves, the loss that is calculated with these $K_3^{1D}$'s is nearly identical.}
%\label{Supplemental2}
%\end{center}
%\end{figure}